\documentclass{article}
\usepackage{spconf,amsmath,graphicx}

\usepackage{cite}
\usepackage{caption}
\usepackage[]{algorithm}
\usepackage{algpseudocode}
\usepackage{todonotes}

\usepackage{amssymb}
\usepackage{verbatim}
\usepackage{hyperref}

\name{Maayan Gelboim$^{\star}$ \qquad Amir Adler$^{\star}$ \qquad Yen Sun$^{\dagger}$ \qquad Mauricio Araya-Polo$^{\dagger}$}
  
  \address{$^{\star}$ Braude College of Engineering, Israel  \\
      $^{\dagger}$TotalEnergies EP Research \& Technology, USA}
\begin{document}

%
\title{Encoder-Decoder Architecture for 3D Seismic Inversion}
\maketitle
\begin{abstract}
Inverting seismic data to build 3D geological structures is a challenging task due to the overwhelming amount of acquired seismic data, and the very-high computational load due to iterative numerical solutions of the wave equation, as required by industry-standard tools such as Full Waveform Inversion (FWI). 
For example, in an area with surface dimensions of 4.5km $\times$ 4.5km, hundreds of seismic shot-gather cubes are required for 3D model reconstruction, leading to Terabytes of recorded data. This paper presents a deep learning solution for the reconstruction of realistic 3D models in the presence of field noise recorded in seismic surveys. We implement and analyze a convolutional encoder-decoder architecture that efficiently processes the entire collection of hundreds of seismic shot-gather cubes. The proposed solution demonstrates that realistic 3D models can be reconstructed with a structural similarity index measure (SSIM) of 0.8554 (out of 1.0) in the presence of field noise at 10dB signal-to-noise ratio.
\end{abstract}

\section{Introduction}
\vskip -5pt
A key step into understanding the subsurface by remote sensing, is the acquisition of seismic data, which consists of the recorded response of the subsurface when mechanical perturbations are introduced.
After data has been collected, several disciplines of geoscience are involved towards the common objective of produce a reliable subsurface model(s). Earth models can be used for many purposes, such as: seismology studies, hydrocarbon exploration and CO$_2$ sequestration. When used for the later purpose, models are critical inputs to drilling decisions. The problem at hand is daunting, too many variables, huge datasets. An example of great societal importance is injecting CO$_2$ from industrial processes into specially reconditioned reservoirs, to that end having high quality subsurface models is crucial. The solution of 3D seismic inverse problems using deep learning (DL) \cite{Goodfellow-et-al-2016,7949028} is an emerging field of research, motivated by state-of-the-art results obtained by DL for the 2D case \cite{9363496,Yang19,Das_2019_impedance}. The contribution of this paper are three-fold: (1) A convolutional encoder-decoder network is proposed with efficient input data dimensionality reduction, to reconstruct complex 3D models at an average inference time of 0.165 seconds (on one NVIDIA A100 GPU), which is a fraction of any iterative global optimization solver.
(2) The proposed approach is demonstrated to provide inherent robustness against noise in the recorded seismic data. (3) The proposed approach is evaluated with realistic 3D geological models. 
\vskip -.2in
\section{Problem Formulation}
\vskip -5pt
Direct reconstruction of models of solid earth is not possible, this renders the following forward model:
\vskip -5pt
\begin{equation}
\textbf{d}=\textbf{F}(\textbf{m})+\epsilon,    
\end{equation} 
only practical when synthetic seismic data ($\textbf{d}$) is to be generated from a forward operator $\textbf{F}$ acting on a artificial model $\textbf{m}$, under ambient noise $\epsilon$. $\textbf{F}$ approximates the behavior of seismic waves propagating through the mechanical medium ($\textbf{m}$), and it is represented by the following expression:
\vskip -5pt
\begin{equation}
\frac{\partial^{2}\mathbf{u}}{\partial t^2} - \mathbf{V_{P}}^{2}\nabla(\nabla \cdot \mathbf{u}) - \mathbf{V_{S}}^{2}\Delta\mathbf{u} = \mathbf{f},
\end{equation}
where $\mathbf{u}=\mathbf{u}(x,y,z,t)$ is the seismic wave displacement, $\mathbf{V_P}$ is P-wave velocity (compression/rarefaction), $\mathbf{V_S}$ is S-wave velocity (shear stress), and $\mathbf{f}$ is the source function.
While the elastic\footnote{In the presence of P-wave and S-wave attenuation the viscoelastic wave equation is utilized instead of the elastic equation.} wave equation describes faithfully seismic waves propagation, it is often preferred (as in this work) to approximate it by the acoustic wave equation \cite{tarantola_inversion_1984}, which assumes only P-waves and requires less computational resources and parameters, as compared to solving the elastic equation. The acoustic wave equation for a medium without density variations is given by:
\begin{equation}
    \frac{\partial^2 \mathbf{u}}{\partial t^2} - \mathbf{V}^2\nabla^2 \mathbf{u} = \mathbf{f},
    \label{eq:acoustic_wave_eq}
\end{equation}
where $\mathbf{u}$ is the wave displacement, $\mathbf{V}$ is the P-wave velocity model and $\mathbf{f}$ is the perturbation source (i.e. \textit{shot}) function. 

Since the direct formulation is not tractable, it is common to use the inverse approach. Seismic velocity inversion computes a complete 3D velocity model ($\hat{\textbf{m}}$) of a certain target area, from recorded seismic data $\mathbf{d_r}$, and it can summarize as:
\vskip -5pt
\begin{equation}
\label{eq:seismic_inverse}
\hat{\textbf{m}}=\textbf{F}^{-1}(\mathbf{d_r}),    
\end{equation} where $\textbf{F}^{-1}$ is the inversion operator. Seismic inversion problems \cite{Schuster17} are ill-posed, e.g., the solution is non-unique and unstable in the sense that small noise variations may alter the solution significantly. The DL formulation for solving inverse problems is detailed in the next section.

\begin{table}[ht!]
\footnotesize	
\caption{Proposed Encoder-Decoder Architecture}
\centering
\begin{tabular}{|l|l|l|l|}
\hline
Block & Layer & Unit &  Parameters  \\
\hline
Input &  0  & Seismic Cube  & $96 \times 96 \times 224$          \\
\hline
 Enc1 &  1  & Conv3D(32, ($5\times 5\times5$))   &  Dropout(0.2), ReLU \\
      &  2  & Conv3D(32, ($5\times 5\times5$))   &  Dropout(0.2), ReLU \\
      &  3  & MaxPool3d   & Dropout(0.2)      \\\hline
Enc2 & 4-6 & Enc1(64) & \\\hline
Enc3 & 7-9 & Enc1(128) &  \\\hline
Enc4 & 10-12 & Enc1(256) & \\\hline
Enc5 & 13-15 & Enc1(512) & \\\hline
Dec1 &  16  & ConvTrans3d     &         ReLU\\

           &  17  &  Conv3D(256, ($5\times 5\times5$))     &  Dropout(0.2), ReLU \\

           &  18  & Conv3D(256, ($5\times 5\times5$))     &  Dropout(0.2), ReLU \\\hline
Dec2 & 19-22 & Dec1(128) & \\\hline
Dec3 & 23-25 & Dec1(64) & \\\hline
Dec4 & 26-28 & Dec1(32) & \\\hline
Output &  29  & Velocity Model  & $96 \times 96 \times 224$         \\
\hline
\end{tabular}
\label{tab:CNN-ENC-DEC}
\end{table}

\section{Deep Learning Approach}
\subsection{Encoder-Decoder Architecture}
\vskip -5pt
Deep Learning (DL) is a powerful class of data-driven machine learning algorithms, built using Deep Neural Networks (DNNs), which are formed by a hierarchical composition of non-linear functions (layers). The main reason for the success of DL is the ability to train very high capacity networks using very large datasets, often leading to and good \textit{generalization} capabilities in numerous problem domains. Generalization is defined as the ability of an algorithm to perform well on unseen examples. In statistical learning terms an algorithm $\mathcal{A:X\rightarrow Y}$ is learned using a training dataset $\mathcal{S} = \{(x_1, y_1), . . . ,(x_N, y_N)\}$ of size $N$, where $\textbf{x}_i\in\mathcal{X}$ is a data sample (in this work, a seismic shot-gather) and $y_i\in\mathcal{Y}$ is the corresponding label (in this work, a 3D  velocity model). Let $\mathcal{P(X,Y)}$ be the true distribution of the data, then the expected risk is defined by:
$
    \mathcal{R(A)}=E_{x,y \sim \mathcal{P(X,Y)}}[\mathcal{L(A}(x),y)],
$ 
where $\mathcal{L}$ is a loss function that measures the misfit between the algorithm output and the data label. The goal of DL is to find an algorithm $\mathcal{A}$ within a given capacity (i.e. function space) that minimizes the expected risk, however, the expected risk cannot be computed since the true distribution is unavailable. Therefore, the empirical risk is minimized instead: 
$
\mathcal{R_E(A)}=\frac{1}{N}\sum_{i=1}^{N}\mathcal{L(A}(x_i),y_i),    
$
 which approximates the statistical expectation with an empirical mean computed using the training dataset.  
 
 \indent In this work we trained a 3D convolutional encoder-decoder, inspired by the 2D U-Net architecture \cite{UNET}, to learn the mapping from seismic data space to 3D models space (i.e. inversion). 
 The complete network is detailed in Table \ref{tab:CNN-ENC-DEC}, with a total of
99M parameters.

 \subsection{Computational considerations}
 The main challenge in training such a deep convolutional neural network (DCNN) for real-life inversion tasks lies in the demanding GPU RAM size and external storage access requirements due to the large number of input channels and large size of each input channel:  each sample in our training data was composed of $N_x \times N_y=529$ seismic data cubes (i.e. DCNN input channels), where $N_x,N_y$ are the total numbers of shots in the lateral and longitudinal axes, respectively. Therefore, a total storage size of ~42GB per sample (after decimation to dimensions $96 \times 96 \times 224$).  A modest training dataset of 300 samples occupies $\approx1.6TB$ storage size, which requires very high-speed storage access to facilitate DCNN training in reasonable duration. Thus, the problem belongs to a High-performance Computing class \cite{akhiyarov20}. To overcome these challenging requirements, we propose a simple yet highly-effective dimensionality reduction scheme: let $\mathbf{d}(S_x,S_y,R_x,R_y,t)$ denote the 5D tensor that represents a single data sample, i.e. the collection of seismic data cubes (shot-gathers), where $S_x,S_y$ are the indices of the shot position, $R_x,R_y$ are the indices of the receiver position, and $t$ is time. We define the time-boosted and dimensionality-reduced data cube $\bar{\mathbf{d}}$ by spatial averaging along the shots dimensions:
 \vskip -15pt
 \begin{equation}
 \label{eq:data-cube}
 \bar{\mathbf{d}}(R_x,R_y,t)=\frac{b(t)}{N_x\times N_y}\sum_{S_x=1}^{N_x}\sum_{S_y=1}^{N_y}{\mathbf{d}(S_x,S_y,R_x,R_y,t)},
 \end{equation}
  where $b(t)$ is a monotonically-increasing time-boosting function that compensates the attenuation of wave reflections from the lowest geological layers, by amplifying late-arrival time samples. Therefore, $\bar{\mathbf{d}}$ forms a single 3D input channel, thus significantly mitigating  the memory and computational requirements for training and inference of the proposed DCNN. 
 In the next section we describe the performance of the proposed architecture for noiseless seismic data, as well as data contaminated by synthetic and field noise.
\vskip -.2in

\section{Performance evaluation}
\vskip -5pt
\subsection{Data Preparation}
\begin{figure}[ht!]
    \centering
    \includegraphics[height=4.0in,width=\columnwidth]{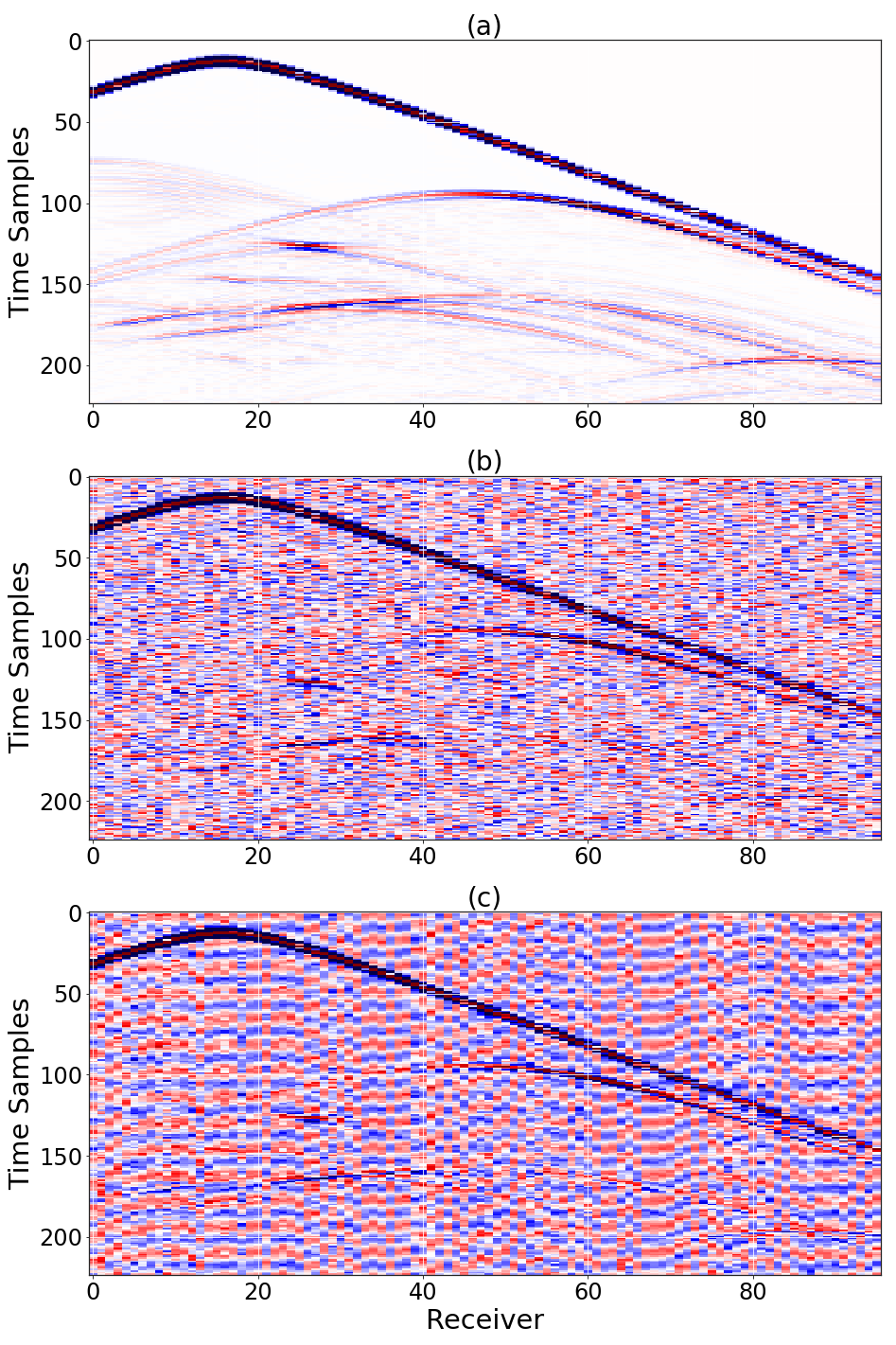}
     \caption{A 2D slice of a 3D shot-gather: (a) noiseless. (b) distorted by additive white Gaussian noise (SNR=10dB). (c) distorted by additive colored field noise (SNR=10dB).}
    \label{fig:noisy_data}
\end{figure}

We created 840 3D velocity models using the Gempy\footnote{\url{https://www.gempy.org/}} tool that creates 3D geologically-feasible models with realistic combinations of features. The selection of Gempy as subsurface modeler is not arbitrary, and obeys to the intention of solving a more realistic problem than just flat layer-cake models. The physical dimensions of each model were $4.5Km \times 4.5Km \times 4.0Km$ (lateral $\times$ longitudinal $\times$ depth),
represented by a 3D tensor of dimensions $300 \times 300 \times 800$ grid points, which was down-sampled for DCNN training to dimensions of $96 \times 96 \times 224$. To generate the synthetic seismic data, through forward modeling, we use an acoustic isotropic wave equation propagator with a 15 Hz peak frequency Ricker wavelet as a source. Shots and receivers are evenly spaced on the top surface of the 3D model (200m between shots and 25m between receivers). To avoid reflections from the boundaries and free surface multiples, convolutional perfectly matched layer (CPML) \cite{li2010convolutional} boundaries are imposed all around the model. Each generated seismic data cube was computed on a grid of dimensions  $180 \times 180 \times 500$ (lateral $\times$ longitudinal $\times$ time) points, which was down-sampled for DCNN training to dimensions of $96 \times 96 \times 224$.
The 840 3D models were split to training and testing sets by a 92\%/8\% ratio, respectively. The proposed DCNN was implemented in PyTorch and trained by minimizing the Mean Absolute Error (MAE) loss function, using the ADAM optimizer with early stopping regularization. Training was repeated for five different scenarios (each resulting in a different trained DCNN): noiseless seismic data, data contaminated with white Gaussian noise at signal-to-noise (SNR) levels of 10dB and 0dB, and data contaminated with field noise (from real data measurements) at SNR levels of 10dB and 0dB. Examples of the clean and noisy data are provided in Fig.~\ref{fig:noisy_data}, demonstrating the highly correlated patterns in space- and time-domains of the field noise.

\begin{figure*}[!htb]
    \centering
    \includegraphics[width=6.0in,height=7in]{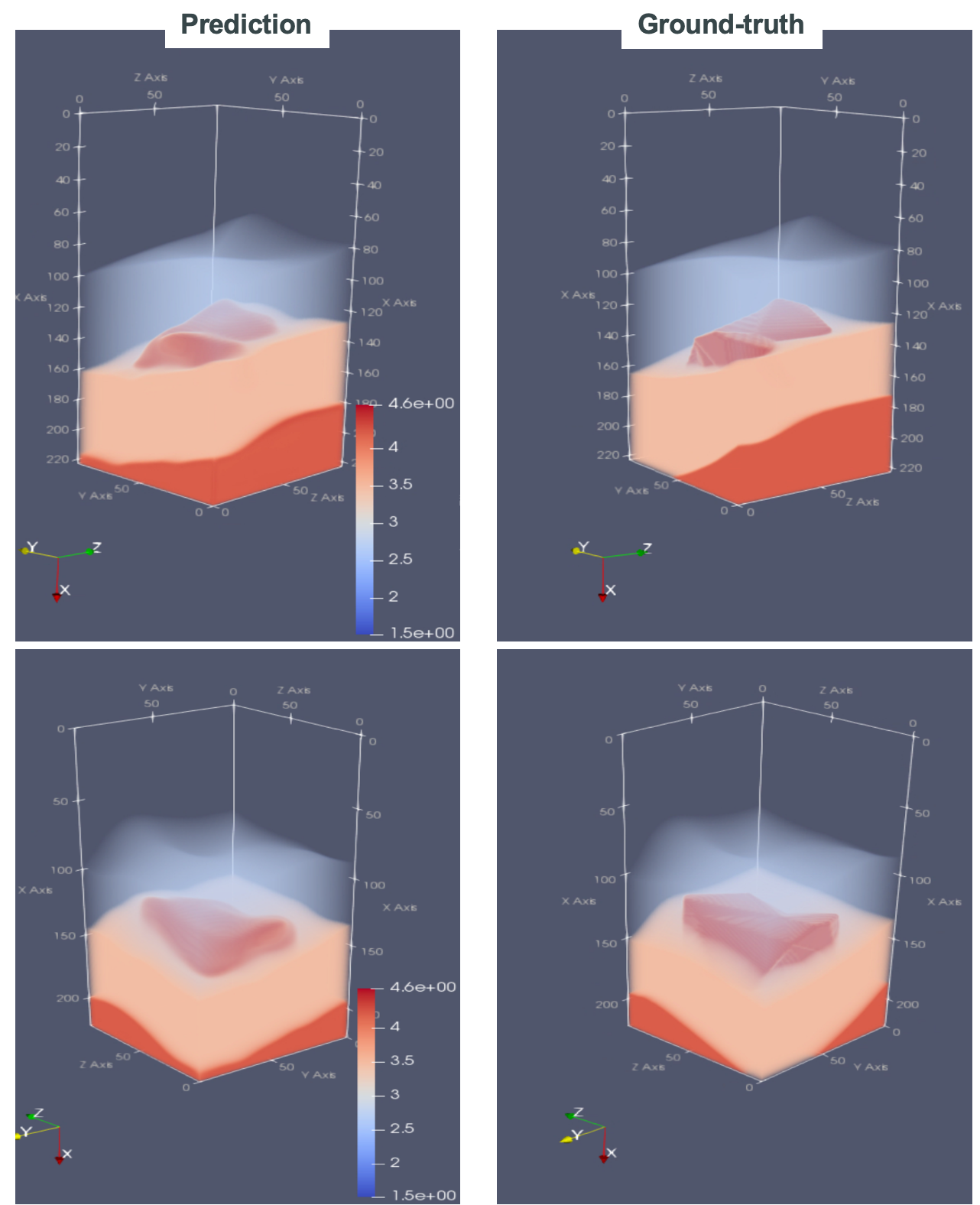}
    \caption{Different angles of a single ground-truth and predicted velocity model, including salt geometry. The shallow layers (slow velocity) are removed to ease the visualization of the embedded salt body.}
    \label{fig:salty}
\end{figure*}

\begin{figure*}[!htb]
    \centering
    \includegraphics[scale=0.5]{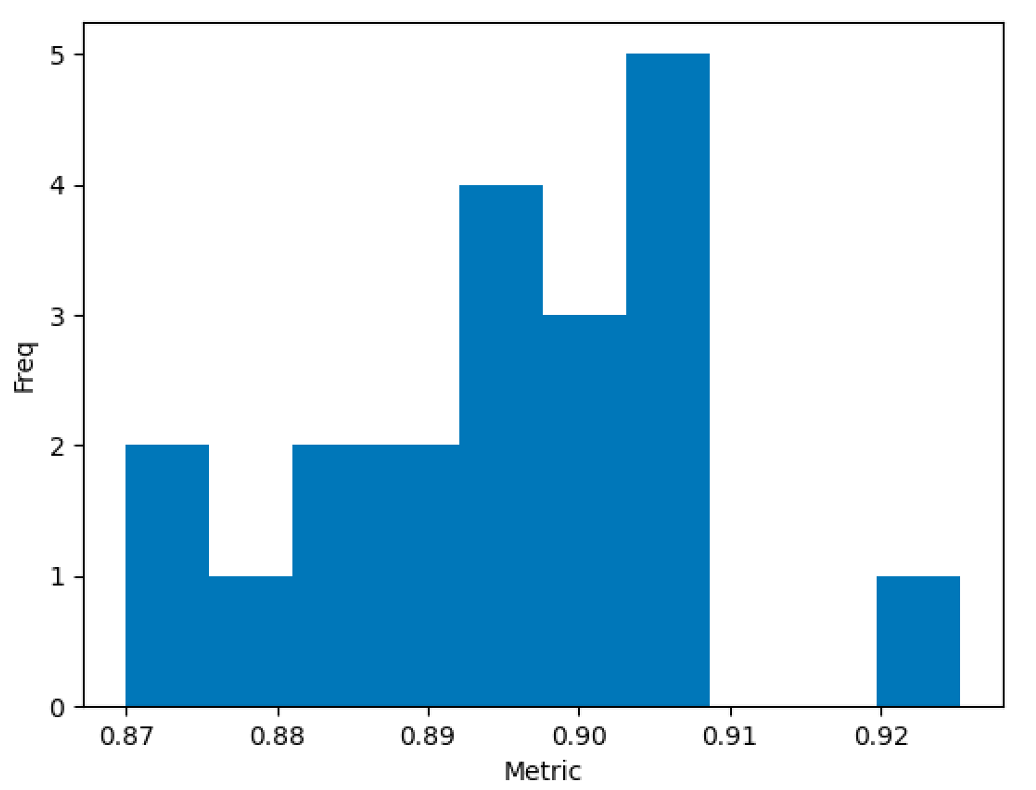}
    \caption{All testing samples SSIM values are within a tight range between 0.87 and 0.92, the average is 0.89.}
    \label{fig:dist}
\end{figure*}

\begin{figure*}[!htb]
    \centering
    \includegraphics[scale=0.5]{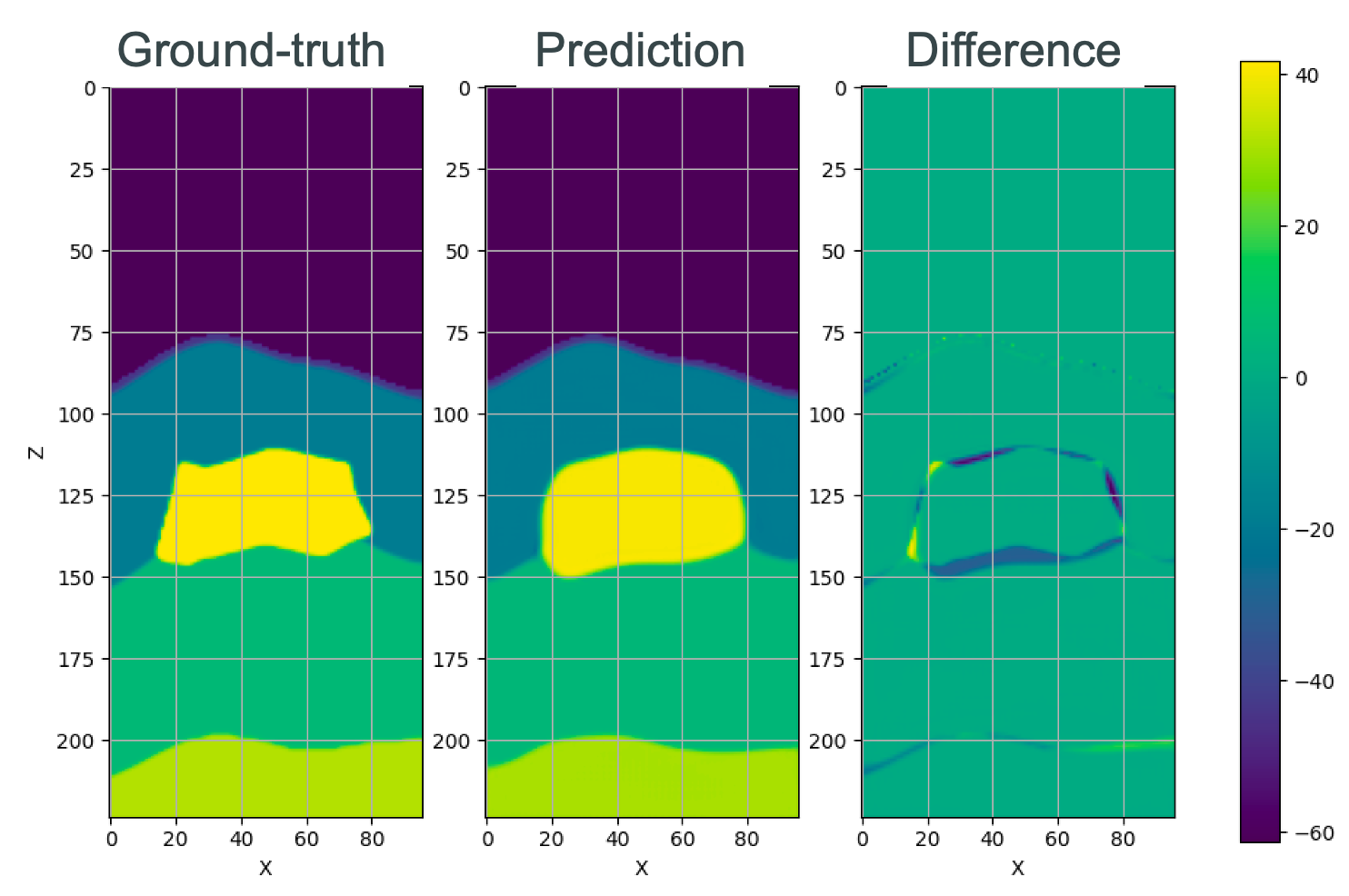}
    \caption{One testing sample is selected and a 2D (central inline) cut is shown, a comparison between ground-truth and prediction is presented in terms of error. As expected, the greater error (right most bar, in percentage) occurs around the salt geometry and mostly over estimating the model velocity, nevertheless the background velocity is overall correct.}
    \label{fig:error}
\end{figure*}

A separated set of 80 samples were augmented with bodies that resemble simple salt geometries. The network trained with 840 samples without noise was used as starting point for a transfer learning process. The resulting transfer learned network was tested on 20 unseen samples also augmented with simple salt geometries. The 3D SSIM for this testing set is 0.89 and the MAE is 55.70 m/s, examples can be observed in Figure~\ref{fig:salty}. The distribution SSIM values for the testing set is presented in Figure~\ref{fig:dist} and an example of the structure of the prediction error (see Figure~\ref{fig:error}).

\subsection{Evaluation Metrics}
\begin{figure*}[!htb]
    \includegraphics[trim={1cm 1cm 1cm 1cm},clip,width=7.0in,height=7in]{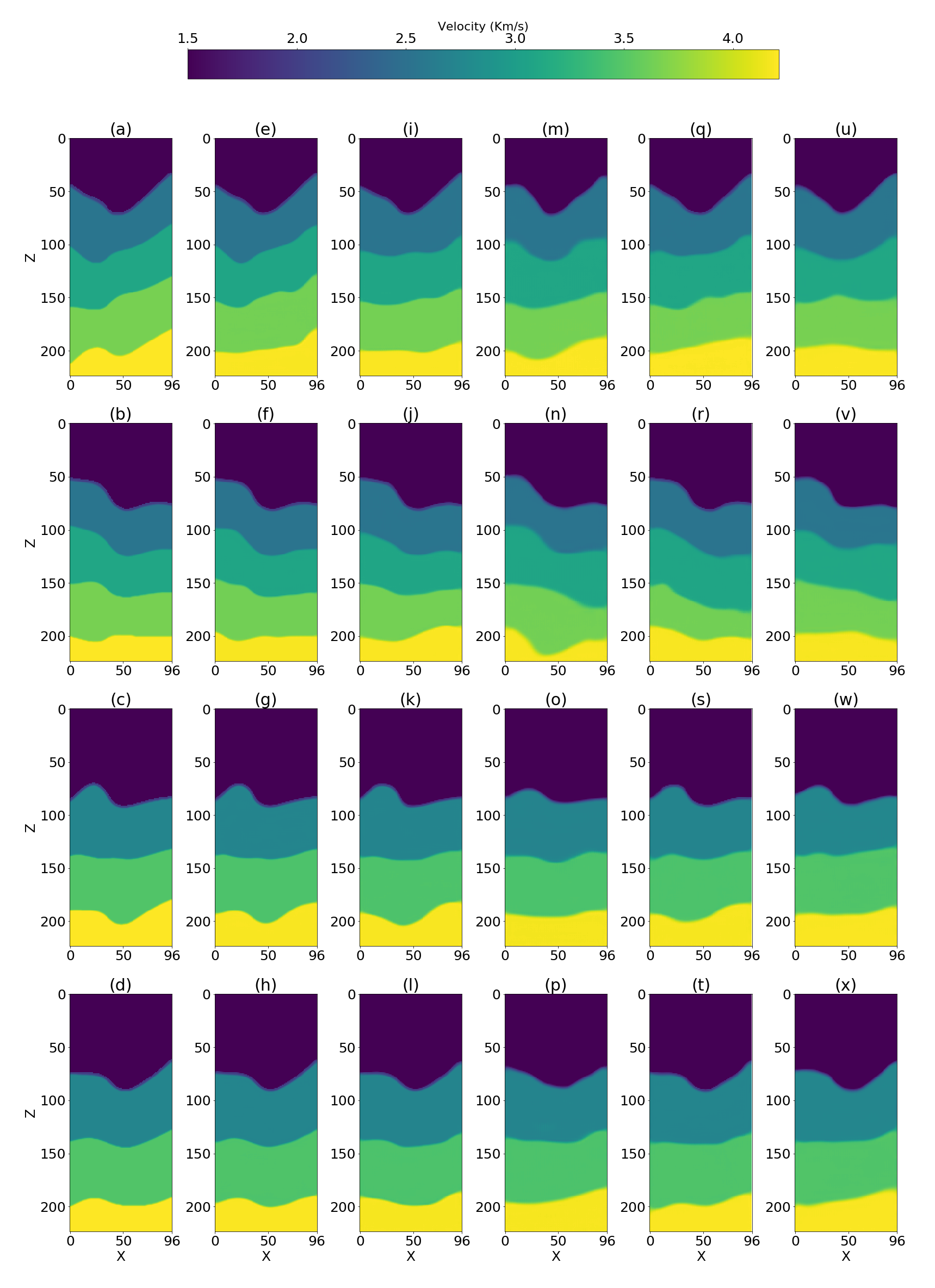}
    \vskip -10pt
    \caption{2D cross-sections of reconstructed 3D models from unseen data: (a)-(d) ground truth. (e)-(h) reconstruction from noiseless data. (i)-(l) reconstruction from noisy data: white noise, SNR=10dB; and (m)-(p) white noise, SNR=0dB. (q)-(t) reconstruction from noisy data: field noise, SNR=10dB; and (u)-(x) field noise, SNR=0dB.}
    \label{fig:my_label}
\end{figure*}

SSIM \cite{1284395} results were computed per 3D model by first averaging SSIM values along the three 2D planes: 96 along the XZ plane, 96 along the YZ plane, and 224 along the XY plane. Finally, the three results were averaged to obtain the single SSIM(3D) result. Table \ref{tab:Results}, details  SSIM and MAE results, averaged on the testing set, clearly demonstrating that the proposed DCNN is capable to reconstruct 3D velocity models from noiseless data (Fig. \ref{fig:my_label}(e)-(h)), as well as with additive white noise (Fig. \ref{fig:my_label}(i)-(p)) or field noise (Fig. \ref{fig:my_label}(q)-(x)). 
Importantly, results  for seismic data contaminated by field noise, indicate close similarity to the ground truth models at a SNR of 10dB, but the SSIM metric slightly deteriorated.

The results in the noiseless data case are surprising, indicating that the dimensionality-reduced data cube (\ref{eq:data-cube}) contains sufficient information for practical reconstruction given the measured metric, achieving an average SSIM(3D) of 0.9003 by the DCNN. We next discuss the noisy data case.

\subsection{Noisy Data Analysis}
In the presence of additive white noise that is spatially (and temporally) independent and identically distributed (iid), the spatial averaging along the shots dimensions results in a reduction of the noise variance, as explained in the following. Denoting by $\mathbf{n}(R_x,R_y,t)$ the noise random variable resulting from the spatial averaging of noise samples,  corresponding to receiver coordinates $(R_x,R_y)$ and time $t$: 
\vskip -15pt
 \begin{equation}
 \label{eq:noisy-data-cube}
 \bar{\mathbf{n}}(R_x,R_y,t)=\frac{1}{N_x\times N_y}\sum_{S_x=1}^{N_x}\sum_{S_y=1}^{N_y}{\mathbf{n}(S_x,S_y,R_x,R_y,t)},
 \end{equation}
where $\mathbf{n}(S_x,S_y,R_x,R_y,t)$ are iid random variables with zero mean and variance  $\sigma^2_\textbf{n}$. By using the iid property, the variance of $\bar{\mathbf{n}}(R_x,R_y,t)$ is independent of $R_x,R_y,t$ and given by
$\sigma^2_{\bar{\mathbf{n}}}(R_x,R_y,t)=\sigma^2_{\bar{\mathbf{n}}}=\frac{1}{N_x\times N_y}\sigma^2_\textbf{n}$ (in our study $N_x\times N_y=529$). The time-boosting function $b(t)$ is omitted from (\ref{eq:noisy-data-cube}), since in the presence of additive noise, both the signal and noise components are multiplied by $b(t)$, according to (\ref{eq:data-cube}), thus the contribution of $b(t)$ is cancelled in a SNR analysis. Therefore, the variance of the noise component in the dimensionality-reduced data cube is effectively reduced by $N_x\times N_y$, for iid white noise. However,  the field noise is clearly not iid, therefore a smaller reduction in the noise variance is achieved. 
\vskip -.05in

\section{Conclusions}
\vskip -.1in
Seismic inversion based on DL effectively reconstruct 3D subsurface models from synthetic seismic data and synthetic seismic data contaminated by either white or field noise. Once training is settled, the inference step is fast - a fraction of a second - to a point that allows many different experiments to be carried out with marginal cost. The robustness to noise also demonstrated to reasonable SNR levels. The next steps for this approach are related to improving the accuracy of the reconstructions for more complex and larger scale structures, and to estimating the sensitivity wrt acquisition towards direct use of field data. 
\vskip -.2in

\section{Acknowledgments}
\vskip -.1in
The authors acknowledge TotalEnergies EP Research \& Technology USA, for supporting this work and allowing its publication.

\begin{table*}[!htb]
\footnotesize
\caption{3D Velocity Model Building Quality Comparison. All values are reported as: Mean(Std), MAE results are in [Km/s].}
\centering
\begin{tabular}{|l|l|l|l|l|l|}
\hline
Data Noise & Noiseless & White (SNR=10dB) & White (SNR=0dB) & Field (SNR=10dB) & Field (SNR=0dB)  \\
\hline
SSIM(3D)  &  0.9003 (0.0679) & 0.86254 (0.06494) & 0.8344 (0.0607) &  0.8554 (0.0599) & 0.8426 (0.0599)      \\
SSIM(XZ)  &  0.9017 (0.0598)  & 0.8586 (0.0528) & 0.8234 (0.0422) & 0.8472 (0.0453) & 0.8313 (0.0408)   \\
SSIM(XY)  &  0.9101 (0.0677) & 0.8875 (0.0657) & 0.8736 (0.0612)  & 0.8875 (0.0606) & 0.8818 (0.0608)  \\
SSIM(YZ)  &  0.8890 (0.0737) & 0.8414 (0.0667) & 0.8060 (0.0555) & 0.8316 (0.0581) & 0.8147 (0.0543)   \\
MAE(3D)   &  
0.0684 (0.0653) & 0.1107 (0.0707) & 0.1499 (0.0677) & 0.1175 (0.0616) & 0.1346 (0.0649) \\
\hline
\hline
\end{tabular}
\label{tab:Results}
\end{table*}

\newpage
\vskip -.2in
\bibliographystyle{IEEEbib}
\bibliography{references}

\end{document}